\documentclass[epsfig,graphics,twocolumn,floatfix,mathbbm,pra]{revtex4}
\usepackage{graphicx}

\begin{document}

\newcommand{\Er}{Er$^{3+}$:Y$_{2}$SiO$_{5}$ }
\newcommand{\levela}{$^{4}$I$_{15/2}$ }
\newcommand{\levelb}{$^{4}$I$_{13/2}$ }
\newcommand{\system}{$\Lambda$-system }

\title{State Preparation by Optical Pumping in Erbium Doped Solids using Stimulated Emission and Spin Mixing}

\author{B. Lauritzen}
\email{bjorn.lauritzen@physics.unige.ch}
\author{S. R. Hastings-Simon}
\author{H. de Riedmatten}
\author{M. Afzelius}
\author{N. Gisin}

\affiliation{Group of Applied Physics, University of Geneva, CH-1211 Geneva 4, Switzerland}

\begin{abstract}

Erbium doped solids are potential candidates for the realization
of a quantum memory for photons at telecommunication wavelengths.
The implementation of quantum memory proposals in rare earth ion
doped solids require spectral tailoring of the inhomogeneous
absorption profile by efficient population transfer between ground
state levels (spin polarization) using optical pumping. In this 
article we investigate the limiting factors
of efficient optical pumping between ground state Zeeman levels in
an erbium doped Y$_{2}$SiO$_{5}$ crystal. We introduce two methods 
to overcome these limiting factors: stimulated emission using a second laser
and spin mixing using radio frequency excitation.  Both methods significantly improve the degree 
of spin polarization.
Population transfer between two Zeeman levels with less than 10
$\%$ of the total population in the initial ground state is
achieved, corresponding to a spin polarization greater than $90\%$. In addition, we demonstrate spectral tailoring by
isolating a narrow absorption peak within a large transparency
window.
\end{abstract}
\date{\today}
\maketitle

\section{Introduction}
In quantum information applications based on optically addressed
atoms or ions it is often necessary to prepare these elements in
specific hyperfine or Zeeman spin levels. Typically this involves
optical pumping from one ground-state spin level to another via an
excited state. The preparation of the ions into a single
well-defined state (spin polarization) is a required initial step for quantum memory
protocols based on electromagnetically induced transparency (EIT)
\cite{Fleischhauer2000, Longdell2005, Eisaman2005,
Chaneliere2005}, Raman interactions \cite{Duan2001,Nunn2007} and
photon echoes using controlled reversible inhomogeneous broadening
(CRIB) \cite{Moiseev2001,Nilsson2005,Kraus2006,Hetet2008} or
atomic frequency combs (AFC) \cite{Afzelius2008}. In particular,
the implementation of these protocols in rare earth (RE) ion doped
solids \cite{Macfarlane2002} requires spectral tailoring of the
inhomogeneous absorption in order to isolate narrow absorption
peaks.

Among the various physical systems that have been considered for
photonic quantum state storage, erbium doped solids provide a
unique system where a large number of stationary atoms can
coherently absorb photons at the telecommunication wavelength of
1.53 $\mu$m. Quantum memories at telecommunication
wavelengths are required for a range of efficient quantum repeater
protocols \cite{Duan2001,Sangouard2007a,Sangouard2008}. In
addition, erbium doped solids have exceptional optical coherence
properties. An optical coherence time as long as 6.4 ms has been
measured in \Er \cite{Sun2002}, which represents the longest optical coherence
time measured in a solid. Another interesting
property of Er$^{3+}$ is that it is a Kramers ion with an odd
number of electrons. This results in a large splitting between the
ground state levels via a first order Zeeman interaction, which 
leads to a larger accessible frequency
bandwidth for quantum memory applications. However, the unquenched
electronic spin of Kramers ions results in strong spin-spin and
spin-phonon interactions as compared to non Kramers ions such as
Pr and Eu. Therefore the ground state population relaxation times are much
shorter than in non Kramers ions, usually in the range of tens to
hundreds of ms \cite{Hastings-Simon2008,Hastings-Simon}. In order
to achieve a high degree of population transfer via optical
pumping it is necessary for the ground state lifetime to be much
longer than the excited state lifetime. Achieving efficient
population transfer in erbium doped materials seems thus
particularly challenging.

The spectroscopic properties of \Er have been extensively studied,
including optical coherence
\cite{Sun2002,Bottger2006,Bottger2006a}, spectral diffusion
\cite{Bottger2006a,Crozatier2007}, hyperfine structure
\cite{Guillot-Noel2006}, Zeeman relaxation lifetimes
\cite{Hastings-Simon}, Zeeman g factors \cite{Sun2008} and
erbium-host interactions \cite{Guillot-Noel2007a}. Slow light has also
been achieved in this material using coherent population
oscillation \cite{Baldit2005}. However, to our knowledge no study
has been reported on the possibility to implement efficient
population transfer between the two Zeeman ground states to 
achieve a high degree of spin polarization.

In this article we investigate optical pumping between ground
state Zeeman levels of erbium ions doped into a Y$_2$SiO$_5$
crystal. We first observe the limitation of standard optical
pumping. For Er$^{3+}$Y$_{2}$SiO$_{5}$ an optical relaxation time
of 11 ms \cite{Bottger2006} and a Zeeman relaxation lifetime of
about 130 ms (at a magnetic field of 1.2 mT) \cite{Hastings-Simon}
have been measured. The low ratio between these two relaxation
lifetimes strongly limits the achievable population transfer
efficiency. Another limiting factor is the branching ratio between
the two optical transitions connecting the two ground state Zeeman
levels. We then show how an enhancement of the optical pumping
efficiency can be achieved by decreasing the excited state
lifetime via optical stimulated emission and by improving spin
branching ratios via radio frequency (RF) excitation. These techniques allow
population transfer between the two Zeeman states with less than
10$\%$ of the total population remaining in the initial state, i.e. 
more than 90$\%$ spin polarization. We also demonstrate spectral
tailoring in this crystal by preparing a
narrow absorption line inside wide transparency window (spectral
pit), as required for the CRIB quantum memory scheme.

\section{Theory}
\label{theory}

Population transfer between the two closely spaced ground state
levels of a \system can be achieved by optical pumping via the
excited state level. In an optical pumping experiment, atoms are
excited by a laser in resonance with the transition connecting one
of the ground state to the excited state. The excited atoms can
then decay into both ground states. Those which have decayed to
the ground state which is not connected to the laser will in
principle remain there for a time corresponding to the relaxation
time between the ground states. If enough pumping cycles can be
done within this time, the population from the initial ground
state can be entirely transferred to the second ground state. The
efficiency of this transfer is thus limited by the ratio of the
excited state lifetime ($T_{1}$) and the spin population lifetime
of the ground state levels (which we label $T_{Z}$, since in our
case the ground state levels are Zeeman levels) as well as the
branching factor $\beta$. The latter is defined as the probability
of the ion to relax into its initial state via spontaneous
emission. 

\begin{figure}[h]
 \centering
\includegraphics[width=\columnwidth]{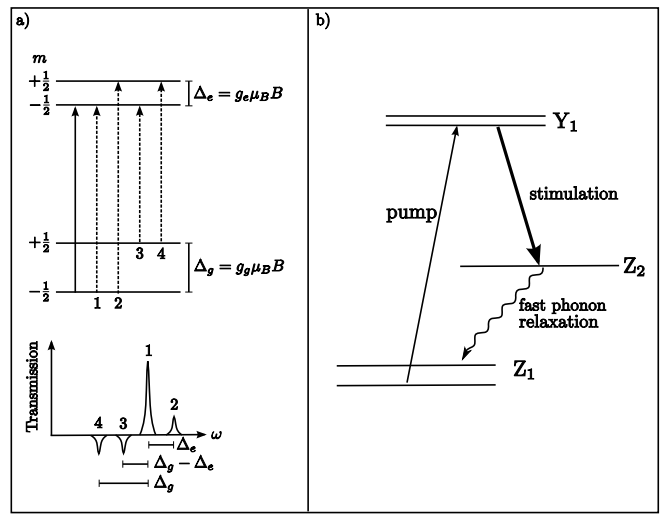}
\caption{a) Spectral holeburning spectrum of a four level system.
The laser is in resonance with the $-1/2 \rightarrow -1/2$
transition (solid line). The probe transitions (dashed lines) are
labelled and the positions of the corresponding holes and
antiholes in the transmission spectrum are shown below. The 
Zeeman splitting of the ground and the excited state
level, respectively, is given by
$\Delta_{e,g}=\mu_{B}g_{e,g}B/\hbar$, where  $g_{g}$
and $g_{e}$ are the g factors for the ground and the excited
state, respectively, $\mu_{B}$ is the Bohr Magneton and $B$ the
magnetic field. Note that in addition to those shown in the
figure, further side- and antiholes occur that are due to the
inhomogeneous broadening of the absorption line
\cite{Hastings-Simon2008,Hastings-Simon,Louchet2007}. b) To
enhance the efficiency of the population transfer the excited
state lifetime is artificially lowered by stimulated emission to
the short-lived second Kramers doublet of the ground state
(Z$_{2}$). } \label{levelscheme}
\end{figure}

In materials with inhomogeneous broadening, population transfer by
optical pumping is usually realized and investigated with spectral
holeburning (SHB) techniques. In this scheme, the pump laser of
frequency $\omega_0$ (in resonance with transition 1, Fig.
\ref{levelscheme}a) is focused on the sample for a duration
$T_{pump}$, the burning time. Ions that are in resonance with
$\omega_0$ are transferred to the excited state. In order to probe
the transmission, a weak pulse follows after a delay $\delta t$
during which the frequency of the laser is scanned around
$\omega_0$ (see Fig. \ref{setup} b). The inhomogeneously broadened
absorption shows a spectral hole at $\omega_0$ which in general
decays with the excited state lifetime $T_{1}$.

Some of the ions will however relax into the second ground
state Zeeman level. If the temperature is low enough ($T<3$ K) such
that Zeeman levels are thermally decoupled \cite{Hastings-Simon},
the population of this level will then be increased and the absorption at
$\omega=\omega_{0}-(\Delta_{g}-\Delta_{e})$ (transition 4 in Fig.
\ref{levelscheme} a) will be enhanced. This enhancement is
called a spectral anti-hole. It will decay with the Zeeman
lifetime $T_{Z}$. As long as the population in the two Zeeman levels
has not relaxed to thermal equilibrium, a
part of the spectral hole at $\omega_{0}$ will persist. In case of
a population transfer the hole  should thus show two different
decay times: $T_{1}$ from ions that decay from the excited back to
the initial state and $T_{Z}$ from ions that relax into the
second Zeeman level. The occurrence of the anti-hole and the long
decay time ($T_{Z}$) for the central hole apparently provide
evidence of population transfer. Throughout this article we speak
of this method as the standard optical pumping scheme.

The intensity of the transmitted light is given by
$I=I_{0}\exp{(-\alpha L)}$, where $I_{0}$ is the intensity of the
incident radiation, and $L$ the length of the sample. The
absorption coefficient $\alpha=\sigma(N_{2}-N_{1})$, i.e. the
logarithm of the transmitted intensity, gives rise to the relation
between the number of ions $N_{1,2}$ in the ground and the excited
state. $\sigma$ is the cross-section of the transition and can be
considered as a constant. Since in this work rather the population
than the measured intensities themselves are of interest, all
curves that deal with quantities related to the absorption and
transmission, respectively, (such as the area or the depth of a
spectral hole) are given with respect to the natural logarithm of
the measured intensity.

To qualitatively understand the limiting factors in the population
transfer process, one can use a simple rate equation model for a
three level $\Lambda$ system. For the steady state in the case of
standard optical pumping (see appendix) it is easy to derive that
the ratio of populations in the ground states after optical
pumping depends on the ratio between the Zeeman lifetime $T_{Z}$
and the excited state lifetime $T_{1}$:

\begin{equation}
 \frac{\rho_{2}}{\rho_{1}}=1+\frac{2T_{Z}}{T_{1}}.
\label{lifetimeratio}
\end{equation}

where $\rho_1$ ($\rho_2$) is the population fraction in the initial
(final) Zeeman state.

However, this holds only for the case in which all of the excited
ions relax into the desired level. In RE-doped solids the
selection rule for the electronic spin $\Delta m=0$ normally is
only slightly lifted by the crystal field and only a small part of
the ions can be found in this decay channel. This motivates to
introduce an effective lifetime in the following manner:
\begin{equation}
 T_{eff}=T_{1}\frac{1}{1-\beta},
\label{efflifetimeratio}
\end{equation}
the branching factor $0\leq\beta\leq1$ being the probability that
an ion will preserve its electron spin upon relaxation from the
excited to the ground state. If $\beta$ is low, the pump rate has
to be accordingly high to transfer population into the excited
state. On the other hand a low branching factor is desirable if
one wants the ions to relax into the second Zeeman level. The
ratio between ions in either of the two ground state levels is then
given by $1+2T_{Z}/T_{eff}$ (see appendix).

In previous measurements on \Er an excited state lifetime of
$T_{1}=11$ms and a branching factor $\beta\geq$0.9 were found
\cite{Hastings-Simon}. This leads to an effective lifetime of
$T_{eff}=110$ ms which is of the same order as $T_{Z}\approx$ 130
ms \cite{Hastings-Simon}. According to the model it is thus
impossible to perform efficient population transfer between the
Zeeman ground states of \Er using standard optical pumping.
However, by looking at Eq.\ref{lifetimeratio} and Eq.
\ref{efflifetimeratio}, one can devise methods to increase the
efficiency of the population transfer. A first idea is to improve
the lifetime ratio by artificially increasing the decay rate from
the excited state. This can be achieved by using stimulated
emission to another short lived ground state level (see
Fig.\ref{levelscheme}b) as proposed in \cite{Gorju2007}. The
corresponding results are presented in section \ref{stimulated}.
Another possibility is to improve the branching ratio by mixing
the spin level in the excited state using a radio frequency (RF)
excitation. This technique is presented in section \ref{spin}.

\section{Experimental Setup}

The crystal, Y$_{2}$SiO$_{5}$ doped with Er$^{3+}$ (10ppm),
belongs to the crystal space group C$_{2h}^{6}$. The Er$^{3+}$
ions replace Y$^{3+}$ ions and occupy two crystallographic
inequivalent sites of C$_{1}$ symmetry (site 1 and site
2) \cite{Bottger2006}.  All measurements in this work were 
performed on ions at
site 1. The relevant transition is from the \levela to the \levelb
level, which are split into eight (Z$_{i}$, $i=1,...,8$ , $i={1}$
labelling the lowest crystal field level) and seven (Y$_{i}$,
$i=1,...,7)$) Kramers doublets, respectively. All measurements were
carried out on the Z$_1\rightarrow$ Y$_1$ transition having a wavelength of 1536 nm.
Under a magnetic field each of the crystal field levels splits into two Zeeman
levels ($m=\pm 1/2)$.

\begin{figure}[h]
 \centering
\includegraphics[width=\columnwidth]{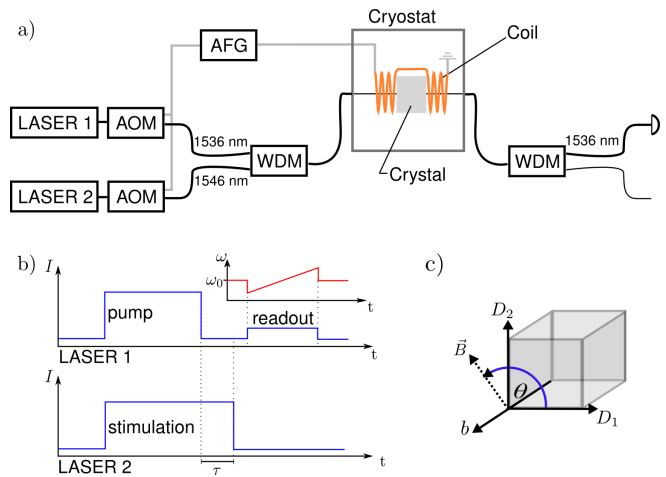}
\caption{a) Experimental setup: The laser for the hole burning ($\lambda=1536$ nm) and
the stimulation laser ($\lambda=1546$ nm) could be intensity modulated independently
with acousto-optic modulators (AOM). They where coupled into the
same fiber via a wave division multiplexer (WDM). The light of the
stimulation laser was filtered out before detection using another
WDM. An RF-wave could be applied on the crystal using a coil in
helmholtz configuration which could be driven with an arbitrary
function generator (AFG). b) Typical pulse sequence. During the
readout pulse the frequency of the pump laser was scanned in order
to record the absorption profile. For measurement with stimulated
emission, the stimulation laser is applied for a time
$T_{pump}+\tau$, where $\tau$ can be varied. c) Illustration of
the angle of the magnetic field with respect to the crystal
orientation} \label{setup}.
\end{figure}

The Y$_{2}$SiO$_{5}$ crystal has three mutually perpendicular
optical-extinction axis labelled $D_{1}$, $D_{2}$, and $b$. The
direction of light propagation is chosen parallel to the $b$-axis.
The crystal was cut along these axes and its dimensions were 3 mm
$\times$ 3.5 mm $\times$ 4 mm along $b$, $D_{1}$, and $D_{2}$,
respectively. The magnetic field was applied in the $D_{1}$-$D_{2}$
plane. In this case all ions of each site are magnetically
equivalent \cite{Sun2008}. The angle of the magnetic field is
defined with respect to the $D_{1}$-axis (see Fig. \ref{setup} c).
If not mentioned explicitly all measurements were taken at an
angle of $\theta=135^{\circ}$.

The experimental setup is shown in figure \ref{setup}a. The
crystal was placed on the cold finger of a liquid helium cryostat
(Janis ST400) and could be cooled down to a temperature of about 2
K. All measurements, if not mentioned differently, were taken at
2.1 K. For optical pumping we used an external cavity diode laser
(Toptica, $\lambda=1536$ nm) which was operated in free run mode and its jitter 
within the relevant timerange was $\geq$ 1 MHz. 
For stimulated emission experiments a tunable diode
laser (Nettest, $\lambda=1546$ nm) was at our disposal. It was amplified with an
erbium doped fiber amplifier (EDFA) and could be tuned over a
wide range. Both lasers could be gated independently using
acousto-optical modulators (AOM) and arbritrary function generators 
(AFG). They were coupled into the same optical fiber
via a wavelength division multiplexer (WDM). The light was focused
into the crystal (diameter of focus ($e^{-2}$) $\approx 70 \mu$m).
In order to avoid a large background in the detection
due to fluorescence from the EDFA the stimulation laser
was filtered out after passing the cryostat using a second WDM.
The transmitted light was measured with a photodiode (NewFocus
2011). Furthermore a Helmholtz coil placed around the crystal was
at our disposal and could be used to apply RF-pulses to the
sample.

\section{Experiments}
\subsection{Standard Optical Pumping}
\label{standard}
\begin{figure}[h]
\centering
\includegraphics[width=\columnwidth]{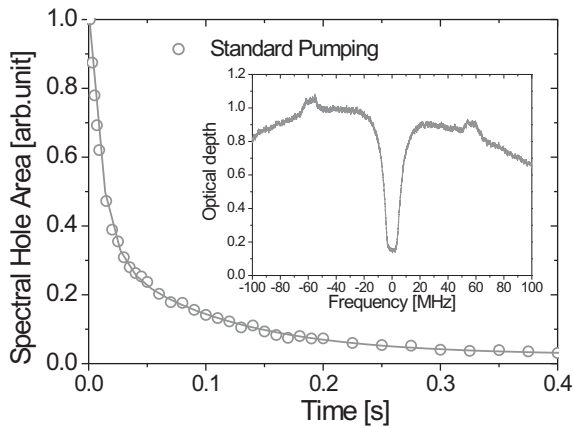}
\caption{Spectral hole area as a function of the readout delay
time for the case of standard optical pumping. The decline is
clearly dominated by the excited state lifetime $T_{1}$ of 11 ms.
The line is a fit to the data. Only a small fraction of the ions
was transferred into the second Zeeman level and provides a slow
component ($T_{Z}\approx$100 ms) to the decline. The origin of the
slowly decaying part is confirmed by the slight enhancement in the
optical depth (anti-holes) left and right of the wide spectral
hole (pit) in the inset. The graph shows a zoom on the absorption
line at about $2.8$ ms after the burning pulse. Here the frequency
of the pump laser was sweeped by 10 MHz during the burning pulse
in order to widen the hole.} \label{nostimnorf}
\end{figure}

 The first series of measurements was made with
standard optical pumping (c.f. section \ref{theory}) in order to
estimate the original efficiency of the population transfer. A
spectral hole was created by sending a resonant pump
pulse of 200 ms into the crystal. The dynamics of the hole was then measured by
probing it at different delays. The circles in figure
\ref{nostimnorf} show the decay of the spectral hole using
standard optical pumping. The dashed line is a fit to the data
given by the sum of two exponential decays with rates of $1/T_{1}$
and $1/T_{Z}$, respectively.

The decay is clearly dominated by the excited state lifetime
$T_1$=11ms. The contribution of the decay with $T_{Z}$ is low
which leads to the conclusion that only a very small fraction of
the ions could be transferred to the second Zeeman level of
Z$_{1}$. This assumption is confirmed by the hole burning spectrum
shown in the inset of  Fig. \ref{nostimnorf}. For this
measurement, a 10 MHz frequency sweep was applied to the pump
laser, in order to widen the spectral hole. The enhancement of
absorption at $\Delta_{g}-\Delta_{e}\approx\pm60$ MHz 
(Fig. \ref{levelscheme}) is low, in contrast to what one would expect 
in the case of an efficient transfer of population.
These observations show that the created spectral
hole is mostly due to a storage of population in the excited
state. Thus, as expected from the arguments given in section \ref{theory},
optical pumping alone does not suffice to prepare the system 
into the required state.

\subsection{Stimulated Emission}
\label{stimulated}

A first possible solution to improve the optical pumping
efficiency is to artificially increase the decay rate from the
excited state by stimulating the emission to another short lived
ground state level (see Fig. \ref{levelscheme}b) \cite{Gorju2007}.
For this purpose, a  second laser in resonance with the transition
Y$_{1} \rightarrow$ Z$_{2}$  (1546 nm, \cite{Bottger2006}) is applied simultaneously with the
pump laser. 
Due to a strong coupling to 
phonon modes, mostly due to direct phonon emission, the excited 
crystal field levels (Z$_{2}$, Z$_{3}$, ...) have non-radiative decay times 
in the range of nano- or picoseconds. Thus, population from these levels
will immediately relax into  ground state (Z$_{1}$).

\begin{figure}[h]
\centering
\includegraphics[width=\columnwidth]{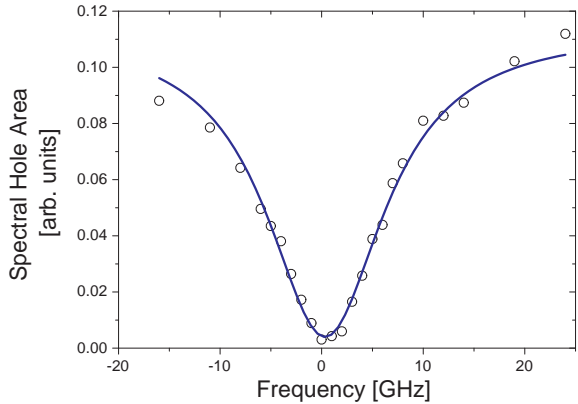}
\caption{(color online) Spectral hole area as a function of the
frequency of the stimulation laser at $T=4.1$ K. The burning time
was 200 ms. The stimulation pulse was 1 ms longer than the burning
pulse. The delay for the readout pulse was set to 2.5 ms. The
solid line is a Lorentzian fit giving a  full width at half maximum (FWHM) of 14 GHz. }
\label{stimfreq}
\end{figure}

We first characterize the stimulation process by measuring its
dependence on the frequency of the stimulation laser. Since
the homogeneous broadening of the transition is expected to be
larger than the inhomogeneous broadening \cite{Bottger2002}, due 
to the short non-radiative decay times, this gives a direct measure 
of the homogeneous linewidth. It has been done 
by measuring the size of the spectral hole as a function of
the frequency of the stimulation laser at 4.1 K. At this
temperature the ground state Zeeman levels are thermally coupled
\cite{Hastings-Simon} and therefore only ions in the excited state
(i.e. ions that have not been stimulated down) contribute to the
hole. The results are shown in Fig. \ref{stimfreq}. For this
measurement, the stimulation pulse was 1 ms longer than the pump
pulse. We see that when the stimulation laser is in resonance, the
spectral hole almost completely vanishes, which shows that one can
efficiently empty the excited state using this method. We find the
linewidth to be about 14 GHz, which corresponds to a lifetime in
the picosecond range. This is 9 orders of magnitude shorter
than the Zeeman lifetime and confirms that the Z$_{2}$-level is
suitable for the  application described above.

\begin{figure}[h]
 \centering
 \includegraphics[width=9cm]{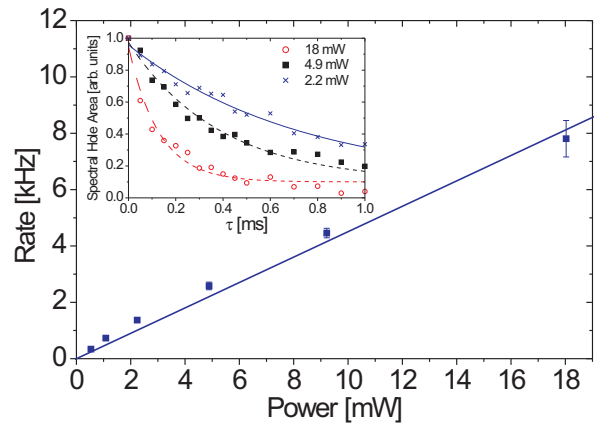}
\caption{(color online) Stimulation rate on the Y$_{1} \rightarrow$ Z$_{2}$ transition vs. applied laser power. The
rate at which ions are stimulated down grows linearly with the
power of the stimulation laser. It was extracted from the decay
curves shown in the inset and as described in the text. Curves for
three different values of the  power of the stimulation laser are
shown. Each data set is normalized to its maximum value
($\tau=0$). The offset of 0.1 is due to persistent holes which
have already been observed and discussed in
\cite{Hastings-Simon}.} \label{stimulation}
\end{figure}

Another quantity of interest is the rate at which stimulation
laser moves the ions down from the excited to the ground state
(Z$_{2}$). To measure this quantity a pump pulse with $T_{pump}=$
100ms was sent into the sample. The crystal temperature was chosen
to be 4.3 K so that also in this case the size of the hole would
only be given by the fraction of ions that remained in the excited
state. Simultaneously to the pump pulse a stimulation pulse with a
duration $T_{pump}+\tau$ was applied (see figure \ref{setup}b).
The size of the remaining hole as a function of $\tau$ was
recorded and the data was fitted with an exponential decay of
which the stimulation rate could be extracted. The inset of Fig.
\ref{stimulation} shows this data for three different values of
applied stimulation power. Each of the lines is a fit to the
respective data. Note that there is an offset which has its origin
in long-lived holes that occur even at this temperature for a
small subset of ions \cite{Hastings-Simon}. It was set to be the
same for all curves. The stimulation rate as a function of the
stimulation power is plotted in Fig. \ref{stimulation}. As one
would expect the stimulation rate increases linearly with power.
For this measurement, the maximal power of the stimulation laser 
on the sample was 20 mW and it was limited by the maximal output 
power of the EDFA. The measurements were taken
with a magnetic field of 27,4 mT at an angle of $\theta=30^{\circ}$ with
the $D_{1}$-axis. The average beam diameter ($e^{-2}$) in the
crystal was 83 $\mu m$.

\begin{figure}[h]
 \centering
 \includegraphics[width=\columnwidth]{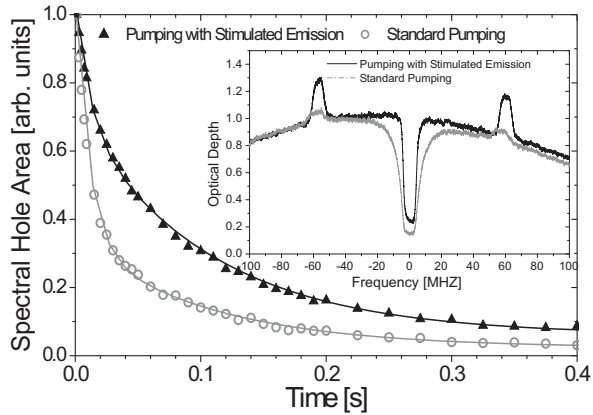}
\caption{ Spectral hole area as a function of the readout delay
time with (triangles) and without (open circles) stimulated
emission. In the case in which the stimulation laser is applied a
clearly enhanced contribution from ions transferred to the second
Zeeman level can be observed. The corresponding hole decays with
$T_{Z}\approx100$ms and the overall decay shows a much stronger
contribution of this slow component than it can be observed for
standard optical pumping. The enhancement in population transfer
is confirmed by the strong anti-holes shown in the inset. The
stimulation power The figure shows the absorption at about
$2.8\pm0.3$ ms after the burning pulse. For this measurement, a 10
MHz sweep was applied on the pump laser and the stimulation pulse
was 1 ms longer than the pump pulse and the stimulation power in
front of the cryostat was 50 mW.} \label{application}
\end{figure}

Let us now investigate the effect of stimulated emission on the
population transfer by optical pumping. In Fig. \ref{application}
one can see that the application of the stimulation laser
simultaneously with the pump laser indeed leads to a significant
enhancement of the transfer efficiency. The decay of the spectral
hole at a temperature of 2.1 K is shown for the case of standard
optical pumping (open circles) and optical pumping assisted by 
stimulated emission (plain triangles). For standard
optical pumping, the decay is dominated by the time constant
$T_{1}$=11ms as explained in Section \ref{standard}. In contrast,
the decay for the stimulated optical pumping is dominated by a
slower decay time, which corresponds to $T_{Z}\approx100$ ms. This
indicates that a significant part of the population is now
transferred to the other Zeeman level by the optical pumping
process.

This hypothesis is confirmed by directly measuring the residual
absorption  after the optical pumping, as shown in the inset of
Fig. \ref{application}. In order to obtain a reference for the
transmission out of resonance, the entire inhomogeneously
broadened absorption line was measured (only a zoom is shown). A
spectral pit was created using stimulated optical pumping by
sweeping the frequency of the pump laser over a range of 10 MHz
while burning. For this measurement, the stimulation pulse was 1 ms
longer than the pump pulse in order to de-excite population remaining 
 in the excited state to the ground state. Therefore the transparency in the
central pit is mostly due to ions transferred to the other ground
state. The optical depth  without optical pumping is $\alpha
L_0=$1, while the residual optical depth after population transfer
is $\alpha L_{res}\approx$ 0.25. The enhanced absorption which
occurs in form of wide anti-holes on both sides of the pit
confirms, as explained in section \ref{theory}, that the
population was transferred as intended. From this measurement, it
is possible to estimate the efficiency of the population transfer.
The fraction of ions remaining in the initial Zeeman state
$\rho_{1,res}$ can be estimated by: $\rho_{1,res}=\alpha
L_{res}/\alpha L_0\approx$ 0.25. Since initially the two Zeeman
states are equally populated, this means that 12.5 $\%$ of the
total number of atoms is remaining in the initial state, within
the frequency range optically pumped. If one assumes that all
atoms that are not in the initial state have been transferred to
the other Zeeman level, on can compute the ratio $\rho_2
/\rho_1$=7. This is much lower than what can be expected from the
simple model presented in section \ref{theory} (see also
Appendix). This shows that, while the model presented helps to
understand the process of optical pumping, it is too simplified to
give quantitative estimations.

\subsection{Spin Mixing} \label{spin}
Although stimulated optical pumping greatly increases the
population transfer efficiency, the fraction of ions remaining in
the initial ground state is still high. This is  in part due to
the high value of the branching factor $\beta$ for this
transition. In Ref. \cite{Hastings-Simon}, $\beta$ was estimated
to be larger than 0.9 for a magnetic field applied at $\theta=135^\circ$
(see Fig. \ref{setup} c), from comparison between experimental
data and numerical simulations. This means that the spin is conserved to a
high degree during the decay from the excited state. From the
measurement of the last section, this seems to hold even for decay
through the the state Z$_2$, which involves phononic relaxation.
In this section, we present a technique to improve the branching
ratio for optical pumping applications.

An idea to overcome this problem in general is to artificially
change the spin by mixing the population of the two Zeeman levels
of the excited state using RF excitation.
For this purpose a pair of Helmholtz coils \footnote{The dimensions of the 
coil used for the spin mixing experiments were: diameter
$= 4$ mm, length of each coil $\approx$ 3 mm, distance between coils
$\approx$ 3 mm, 3 turns on each side. The central axis was parallel to the axis of light
propagation.} was placed around the crystal (see Fig. \ref{setup}a).
Note that a difference of the g-factors for the ground- and the excited state is
necessary in order not to mix the spin population of the ground
state levels as well. In \Er $g_{g}=12$ and $g_{e}=8$ (for
$\vec{B}$ at an angle of $\theta=135^{\circ}$, see Fig. \ref{setup} c)
\cite{Sun2008}. For electronic spins with such large g-factors, the
magnetic dipole moment of the spin transition is large ($\mu$=112 MHz/mT),
so that significant mixing can be obtained with moderate 
RF-power.  
The RF pulse was applied simultaneously with the
pump pulse (duration 200 ms). A 1 $\mu$s long linear frequency
sweep was continuously repeated during the burning process. Due to
the inhomogeneity of the external magnetic field the bandwidth,
i.e. the size of the sweep, had to be chosen high (10-20 MHz). The
required central frequency corresponds to $\Delta_{e}$ in figure
\ref {levelscheme} and could be extracted from a hole-burning
spectrum.
\begin{center}
    \begin{figure}[h]
    \centering
    \includegraphics[width=\columnwidth]{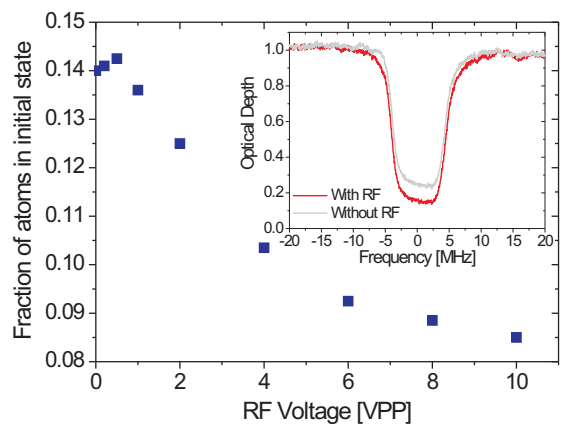}
\caption{Efficiency of the spin-mixing method as a function of the
applied RF-voltage. The central frequency of the RF wave is 135
MHz, corresponding to the splitting of the excited state Zeeman
levels. The efficiency is estimated by measuring the fraction of
the total number of atoms remaining in the initial state after
optical pumping. Both stimulated emission (50 mW) and RF mixing
are used. The duration of the pump sequence is 200 ms. The
stimulation pulse is 1 ms longer than the pump pulse, in order to
empty the excited state. The inset shows an example of a spectral
pit created by population transfer with a frequency sweep of 10
MHz, without and with RF mixing (with the highest RF voltage of 10 Vpp).}
    \label{rf-power}
    \end{figure}
\end{center}
The effect of the RF mixing was measured by estimating the
fraction of atoms remaining in the initial state after optical
pumping, similarly to the measurement made in section
\ref{stimulated}. For this measurement however, both stimulated
emission and RF excitation were applied. Fig. \ref{rf-power} shows
the fraction of the total number of atoms remaining in the initial
state as a function of the RF voltage. A significant enhancement
of the transfer due to the RF-wave can be observed. This
enhancement, however, is quickly saturated when
increasing the RF-power. This might be explained by the
achievement of an equilibrium between the population of the two
Zeeman levels. The inset of Fig.\ref{rf-power} shows a zoom of the
spectral pit created with and without RF mixing. For this
measurement, the number of atoms in the initial state was 8 $\%$, 
as compared to 12.5 $\%$ without RF mixing applied (Sec. \ref{stimulated}). Hence, a spin 
polarization of 92 $\%$ within the optically pumped bandwith has been achieved.

\subsection{Spectral Tailoring}
\begin{center}
    \begin{figure}[h]
    \centering
    \includegraphics[width=\columnwidth]{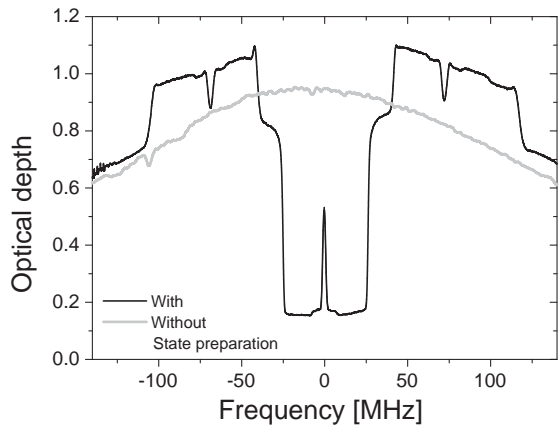}
    \caption{Spectral tailoring in \Er (black curve): A narrow absorption line
(FWHM 2 MHz) could be established inside a spectral pit (FWHM 50
MHz). An interesting feature is that the structure written into
the absorption can as well be observed in the corresponding
anti-holes, which again confirms the population transfer into the
second Zeeman level of the ground state. The gray curve
corresponds to the inhomogeneous absorption without the state
preparation sequence. For further details see text.}
    \label{peak}
    \end{figure}
\end{center}
The ability to tailor the absorption profile is important for
quantum memory proposals such as CRIB or AFC, as well as proposals for 
quantum logic gates in RE doped solids \cite{Rippe2005}. To illustrate this
capacity in \Er, we isolated a narrow absorption line within a
large transparency window (spectral pit), with the help of the two
methods presented above. For this purpose, we used a simple
technique consisting of sweeping the laser frequency during the
200 ms long pump pulse and gating off the pump laser for a short
time at the center of the frequency sweep. The total frequency
range was swept in 100 $\mu$s, resulting in 2000 frequency sweeps. An
example of spectral tailoring is shown in Fig. \ref{peak}. The
width of the pit was 50 MHz while the absorption width of the 
peak at the center of the pit is 2 MHz. 
This is much broader than the homogeneous linewidth of the
transition \cite{Bottger2006}. The width of the peak is likely to be
limited by the linewidth of our free-running laser, which is
larger than 1 MHz for the timescale of the burning pulse. The fact
that we sweep the laser frequency by scanning the laser current
might also degrade the linewidth. Stabilizing the laser frequency
will be required in order to obtain narrower absorption peaks.
Furthermore, the use of more sophisticated hole burning techniques
will also be needed \cite{Pryde2000,Sellars2000,Seze2003,
Nilsson2004,Rippe2005} in order to achieve absorption width in the
kHz range. Although the methods developed in this paper allow a
significant improvement of population transfer efficiency as
compared to standard optical pumping, there is still a residual
absorption in the spectral pit. This will act as a passive loss
for light storage protocols. Further efforts will thus be needed
in order to optimize light storage efficiency. Note however that
the width of the spectral pit (50 MHz) is much larger than what
can be achieved in non-Kramers ions that have been used for such 
purposes so far, such as praesodymium \cite{Hetet2008,Nilsson2004} or
europium \cite{Nilsson2002,Pryde2000}. This shows the potential of Er doped solids for
obtaining large frequency bandwidths for quantum storage
applications.

\section{Conclusions}
Erbium doped solids are promising candidates for the
implementation of quantum memories at telecommunication
wavelengths. We have shown and discussed the difficulties for
state preparation by optical pumping in \Er, including the ratio
between the excited state and the Zeeman lifetimes and the
branching factor $\beta$. With stimulated emission assisted
optical pumping and spin-mixing using RF excitations, we have
investigated two methods to alleviate these problems. We have
shown that their application indeed leads to a significant
enhancement of the optical pumping efficiency. Population transfer
between two Zeeman levels with less than 10$\%$ of the total
population in the initial level has been demonstrated. This corresponds 
to a spin polarization of 90$\%$.  Note that
the techniques used here can also be applied to other rare earth
doped solids. In particular, the might be useful to enhance
optical pumping efficiency in other Kramers ions, such as
neodymium \cite{Hastings-Simon2008}.

Further improvements might be obtained by increasing stimulation
rates with higher laser powers and using coherent RF population
transfer between the two excited states. Another possibility to
improve the branching ratio could be to apply the magnetic field
at particular angles where the spin conservation selection rule is
relaxed \cite{Seze2006,Louchet2007,Thiel}. In order to increase
the population lifetime of Zeeman levels, it might be interesting
to work at higher magnetic field (hundreds of mT), in order to
reduce spin-spin interactions. Other crystals might also be
explored, e.g. Y$_2$O$_3$, to search for longer Zeeman lifetimes.
Finally, it would also be interesting to investigate hyperfine
states, which might have longer spin population relaxation
lifetimes \cite{Guillot-Noel2006}.

We also demonstrated spectral tailoring in \Er by isolating a
narrow absorption peak within a wide transparency window, as
required for the CRIB quantum memory scheme. Although further
progress remains to be done to improve the population transfer
efficiency with higher optical depth, the achievement of spectral
tailoring in erbium doped solids will allow proof of principle
experiments of light storage with quantum memory protocols. Hence,
with this work a further step towards a solid state based quantum
memory at telecom wavelength has been accomplished.

\section{Acknowledgments}
We would like to thank Marko Lovric and Roman Kolesov for their
advice concerning the design of the RF-setup. Furthermore would we
like to thank the group of Prof. D. van der Marel, Mehdi Brandt,
D. van Mechelen, and Violeta Guritanu for experimental assistance,
as well as Prof. M. Chergui (EPFL) for loaning us the cryostat.
Tecnical support by C. Barreiro and J.D. Gauthier is acknowledged.
This work was supported by the Swiss NCCR Quantum Photonics and
the European Commission under the Integrated Project Qubit
Applications (QAP) funded by the IST directorate.

\appendix
\section{Lifetime Ratio}
\subsection{Standard optical pumping}
The low efficiency of the population transfer was ascribed to the
the ratio between the excited state lifetime $T_{1}$ and the
ground state Zeeman lifetime $T_{Z}$ as well as the branching
factor $\beta$. The corresponding formulas (\ref{lifetimeratio}
and \ref{efflifetimeratio}, section \ref{theory})  were found
using the following simple rate equation model:
\begin{figure}[h]
\centering
\includegraphics[width=80pt]{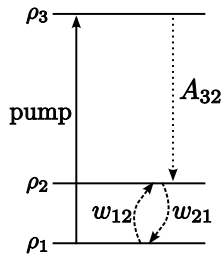}
\caption{Levelscheme to estimate the population ratio in the case
of standard optical pumping.} \label{popratio}
\end{figure}

Let $\rho_{1}$, $\rho_{2}$ and $\rho_{3}$ be the population fraction in
the two ground and in the excited state, respectively, as shown in figure
\ref{popratio}. The laser is in resonance with the transition
between level 1 and 3. For the population of the second Zeeman
ground state level the population dynamics is given by
\begin{equation}
\frac{d
}{dt}\rho_{2}=w_{12}\rho_{1}-w_{21}\rho_{2}+A_{32}\rho_{3}.
\label{dynamics}
\end{equation}
$A_{32}$ is the Einstein-coefficient for spontaneous emission
$w_{12}$ and $w_{21}$ are the spin-lattice-relaxation rates. In
the case considered by Gorju et al. \cite{Gorju2007} a term which
corresponds to $w_{12}$ does not occur since the target state is a
crystal field level and the relevant phonon decay is a one way
process. However, the calculations are still quite similar. In the
steady state it is
\begin{eqnarray}
\frac{d }{dt}\rho_{2} &= & 0 \\
\rho_{1}& = &\rho_{3},
\end{eqnarray}

where the second equality holds for the case of sufficiently high pump power, i.e. a saturation of the considered transition.
If we further assume $w_{12}=w_{21}$, which is a good approximation since the thermal energy $k_{B}T$ is much larger than the splitting between the two levels, it follows from equation
\ref{dynamics} that the ratio between the number of ions in either
of the lower states is given by
\begin{equation}
\frac{\rho_{2}}{\rho_{1}}=1+\frac{A_{32}}{w_{12}}=1+\frac{2T_{Z}}{T_{1}}.
\label{ratio1app}
\end{equation}
The factor 2 in the numerator of the lifetime ratio occurs since
$T_{Z}=1/(w_{12}+w_{21})=1/(2w_{12})$ \cite{Geschwind1972}.

So far it is assumed that all ions decay into the second Zeeman
level. In this case $1/A_{32}=T_{1}$. However, one has to take
into account that only a fraction of the ions in the excited state
will do so. Supposing that the excited state lifetime is given by
$T_{1}=1/A$ with $A=A_{31}+A_{32}$ and with the definition of the
branching factor $\beta=A_{31}/(A_{31}+A_{32})$ (which gives
$1-\beta=A_{32}/(A_{31}+A_{32})$) we can define the effective
lifetime
\begin{equation}
T_{eff}=\frac{1}{A_{32}}=\frac{T_{1}}{1-\beta}.
\end{equation}
Plugging this into equation \ref{ratio1app} we get
\begin{equation}
\frac{\rho_{2}}{\rho_{1}}=1+\frac{2T_{Z}}{T_{eff}}.
\label{ratio2app}
\end{equation}

\subsection{Stimulated emission} In the case of stimulated emission as
presented in section \ref{stimulated} the spontaneous emission
rate $A$ has to be replaced with the stimulation rate $\Gamma$.
Due to its extremely low lifetime, the role of the level Z$_{2}$
(figure \ref{levelscheme}) can be neglected. According to equation
\ref{ratio2app}  population ratio is now given by
\begin{equation}
\frac{\rho_{2}}{\rho_{1}}=1+2(1-\beta)\Gamma T_{Z}
\end{equation}
If one considers a branching factor $\beta=0.95$ and a stimulation
rate of $\Gamma=7 $kHz, with a Zeeman lifetime of 100 ms we
compute a ratio of 71, which is significantly higher ($\sim$ one order of
magnitude) than found in the experiment. This can be explained by
the fact that the model is extremely simplified as compared to
reality. Among others the role of other decay channels and the
possibility of a different branching factor for a decay from the
Z$_{2}$ level could not been taken into account. Nevertheless,
this model can help to understand the processes of optical pumping
in the system qualitatively.

\end{document}